\input epsf.tex
\magnification=1200
\hsize=15truecm
\vsize=23truecm
\baselineskip 18 truept
\voffset=-0.5truecm
\parindent=0cm
\overfullrule=0pt
\def\Ai{\hbox{\hbox{${\cal A}$}}\kern-1.9mm{\hbox{${/}$}}}
\def\Vi{\hbox{\hbox{${\cal V}$}}\kern-1.9mm{\hbox{${/}$}}}
\def\Di{\hbox{\hbox{${\cal D}$}}\kern-1.9mm{\hbox{${/}$}}}
\def\lam{\hbox{\hbox{${\lambda}$}}\kern-1.6mm{\hbox{${/}$}}}
\def\D{\hbox{\hbox{${D}$}}\kern-1.9mm{\hbox{${/}$}}}
\def\A{\hbox{\hbox{${A}$}}\kern-1.5mm{\hbox{${/}$}}}
\def\V{\hbox{\hbox{${V}$}}\kern-1.9mm{\hbox{${/}$}}}
\def\parz{\hbox{\hbox{${\partial}$}}\kern-2mm{\hbox{${/}$}}}
\def\B{\hbox{\hbox{${B}$}}\kern-1.7mm{\hbox{${/}$}}}
\def\R{\hbox{\hbox{${R}$}}\kern-1.7mm{\hbox{${/}$}}}
\def\si{\hbox{\hbox{${\xi}$}}\kern-1.7mm{\hbox{${/}$}}}



\def\pmb#1{\leavevmode\setbox0=\hbox{$#1$}
\kern-.025em\copy0\kern-\wd0
\kern-.05em\copy0\kern-\wd0\kern-.025em\raise.0433em\box0}


%
\catcode`@=11
%
%
%
\def\lsim{\mathchoice
  {\mathrel{\lower.8ex\hbox{$\displaystyle\buildrel<\over\sim$}}}
  {\mathrel{\lower.8ex\hbox{$\textstyle\buildrel<\over\sim$}}}
  {\mathrel{\lower.8ex\hbox{$\scriptstyle\buildrel<\over\sim$}}}
  {\mathrel{\lower.8ex\hbox{$\scriptscriptstyle\buildrel<\over\sim$}}} }
\def\gsim{\mathchoice
  {\mathrel{\lower.8ex\hbox{$\displaystyle\buildrel>\over\sim$}}}
  {\mathrel{\lower.8ex\hbox{$\textstyle\buildrel>\over\sim$}}}
  {\mathrel{\lower.8ex\hbox{$\scriptstyle\buildrel>\over\sim$}}}
  {\mathrel{\lower.8ex\hbox{$\scriptscriptstyle\buildrel>\over\sim$}}} }
\def\croce{\displaystyle / \kern-0.2truecm\hbox{$\backslash$}}
\def\lqua{\lower4pt\hbox{\kern5pt\hbox{$\sim$}}\raise1pt
\hbox{\kern-8pt\hbox{$<$}}~}
\def\gqua{\lower4pt\hbox{\kern5pt\hbox{$\sim$}}\raise1pt
\hbox{\kern-8pt\hbox{$>$}}~}
\def\mma{\lower1pt\hbox{\kern5pt\hbox{$\scriptstyle <$}}\raise2pt
\hbox{\kern-7pt\hbox{$\scriptstyle >$}}~}
\def\mmb{\lower1pt\hbox{\kern5pt\hbox{$\scriptstyle >$}}\raise2pt
\hbox{\kern-7pt\hbox{$\scriptstyle <$}}~}
\def\mmc{\lower4pt\hbox{\kern5pt\hbox{$<$}}\raise1pt
\hbox{\kern-8pt\hbox{$>$}}~}
\def\mmd{\lower4pt\hbox{\kern5pt\hbox{$>$}}\raise1pt
\hbox{\kern-8pt\hbox{$<$}}~}
\def\lsu{\raise4pt\hbox{\kern5pt\hbox{$\sim$}}\lower1pt
\hbox{\kern-8pt\hbox{$<$}}~}
\def\gsu{\raise4pt\hbox{\kern5pt\hbox{$\sim$}}\lower1pt
\hbox{\kern-8pt\hbox{$>$}}~}
\def\croce{\displaystyle / \kern-0.2truecm\hbox{$\backslash$}}
\def\ali{\hbox{A \kern-.9em\raise1.7ex\hbox{$\scriptstyle \circ$}}}
\def\2frecce{\hbox{\lower 0.3ex\hbox{$\leftarrow$}
\hbox{\kern-1.3em\raise 0.3ex\hbox{$\rightarrow$}}}}
%
%
%
%
\def\quad@rato#1#2{{\vcenter{\vbox{
        \hrule height#2pt
        \hbox{\vrule width#2pt height#1pt \kern#1pt \vrule width#2pt}
        \hrule height#2pt} }}}
\def\quadratello{\mathchoice
\quad@rato5{.5}\quad@rato5{.5}\quad@rato{3.5}{.35}\quad@rato{2.5}{.25} }
%
%
\font\s@=cmss10\font\s@b=cmbx8
\def\reali{{\hbox{\s@ l\kern-.5mm R}}}
\def\m{{\hbox{\s@ l\kern-.5mm M}}}
\def\k{{\hbox{\s@ l\kern-.5mm K}}}
\def\naturali{{\hbox{\s@ l\kern-.5mm N}}}
\def\interi{{\mathchoice
 {\hbox{\s@ Z\kern-1.5mm Z}}
 {\hbox{\s@ Z\kern-1.5mm Z}}
 {\hbox{{\s@b Z\kern-1.2mm Z}}}
 {\hbox{{\s@b Z\kern-1.2mm Z}}}  }}
\def\complessi{{\hbox{\s@ C\kern-1.7mm\raise.4mm\hbox{\s@b l}\kern.8mm}}}
\def\toro{{\hbox{\s@ T\kern-1.9mm T}}}
\def\unity{{\hbox{\s@ 1\kern-.8mm l}}}
%
%
\font\bold@mit=cmmi10
\def\setbmit{\textfont1=\bold@mit}
\def\bmit#1{\hbox{\textfont1=\bold@mit$#1$}}
%
\catcode`@=12

\null
\vskip 1.5truecm

\centerline{\bf U(1) $\times$ SU (2) GAUGE THEORY OF}
\centerline{\bf UNDERDOPED HIGH $T_c$ CUPRATES VIA}
\centerline{\bf CHERN--SIMONS BOSONIZATION}

\vskip 1truecm
\centerline{P.A. MARCHETTI}

\vskip 0.5truecm
\centerline{\sl Dipartimento di Fisica, Universit\`a  di Padova}

\centerline{\sl and}

\centerline{\sl INFN -- Sezione di Padova, I--35131 Padova, Italy}

\vskip 2truecm
\centerline{\bf Abstract}
\vskip 0.3 truecm
We outline the basic ideas involved in a recently proposed [17] derivation of
a gauge theory for underdoped cuprates in the ``spin--gap phase",
performed essentially step by step starting from the $t-J$ model,
considered as a model Hamiltonian for the $CuO_2$ layers. The basic tool is
the $U(1)\times SU(2)$ Chern--Simons bosonization, to which  it is
dedicated
a somewhat detailed discussion. The basic output is a ``spin--gap"
not vanishing in any direction and an antiferromagnetic correlation
length proportional to the inverse square root of doping concentration,
in agreement with data deduced from the neutron experiments. The model
also exhibits a small half--pocket
Fermi surface around $(\pm \pi/2,\pm \pi/2)$ and a linear in temperature
dependence of in--plane resistivity in certain temperature range.

\vfill\eject

{\sl 1. High $T_c$ Cuprates and the t--J model}
\vskip 0.3truecm

A common structural feature of high $T_c$ cuprates is the presence of
electronically active $Cu O_2$ layers, alternating with (insulating) block
layers along the crystalline  $c$--axis.
In the $Cu O_2$ layers of undoped materials the 3$d$ shell of copper has
a hole (primarily in the highest energy $3d_{x^2 - y^2}$ orbital) while the
2$p$
shell of the oxygen is filled. The spin ${1\over 2}$ moments of the $Cu$ are
antiferromagnetically ordered at low temperatures and a strong on--site
Coulomb repulsion acts in the $3d$ orbital, inhibiting double
occupation.
As the materials are doped, holes (or electrons) are introduced in the
$Cu O_2$ layers.

In terms of doping concentration ($\delta)$ and temperature ($T$), a
``typical" phase diagram is drawn in fig.1 (patterned on
$La_{2-\delta} Sr_\delta
Cu O_4$ compounds). It exhibits an antiferromagnetic (AF) insulating phase
near $\delta =0$ (for sufficiently low temperatures), a superconducting (SC)
phase for an intermediate doping (e.g. 0.1 $\lqua\delta \lqua 0.2$ for $La
Sr Cu O$ compounds). The materials with doping concentration exhibiting
highest $T_c$ are called optimally doped; cuprates with lower or higher
doping concentration are called underdoped or overdoped, respectively.
At optimal doping, in
the phase diagram  above the SC region there is a region
characterized by an anomalous metallic behaviour (e.g. linear in $T$ in--plane
resistivity $\rho_{ab}$ [1]; anomalous spin lattice relaxation rate ${1\over
T_1 T} \sim {1\over T^\alpha}, \alpha \sim 1$ [2]; large 2D Fermi surface
whose volume is consistent with the Luttinger theorem [3]). Moving towards the
underdoped region there is a crossover to the so--called ``spin--gap phase",
exhibiting distinctive phenomena (e.g. a minimum of
$\rho_{ab} (T)$ for small
enough $\delta$ [4]; a maximum in ${1\over T_1 T}$ at low $T$ [5];
in some materials small
half--pocket like 2D Fermi surface around $(\pm {\pi\over 2}, \pm {\pi\over
2})$[6]).
In the overdoped region the materials appear to show an essentially
``normal" metallic behaviour.

In this paper we will be interested in the
underdoped (spin--gap) region. A model Hamiltonian for the $Cu O_2$ planes
(at low doping) has been proposed by Zhang and Rice  [7], following a
suggestion
by Anderson [8], roughly on the basis of the following considerations (see
[9] for
a more precise and detailed discussion). The holes introduced by doping
go primarily into symmetrized $O$--orbitals around the $Cu$ ion and they
form a
spin singlet with the spin moment of copper (see fig. 2). A spin singlet
($Cu$--hole /$O$--hole) in one $Cu O_4$ has a relevant nearest neighbour (n.n.)
hopping, since each $Cu O_4$ has an $O$--site in common with the n.n.
$Cu O_4$.

The low energy physics is then believed to be dominated by the motion of
these spin--singlets in the AF background of $Cu$ spin moments. The
Hamiltonian proposed for the system is given by

$$
H = P_G [\sum_{<ij>} \sum_\alpha - t (c^\dagger_{i\alpha} c_{j\alpha} +
h.c.) + J \vec S_i \cdot \vec S_j] P_G, \eqno(1.1)
$$

where $i$ runs over the sites of the (square) lattice defined by the
position of the $Cu$ ions, the sum over $\alpha$ runs over spin indices
(spin up=1, spin down=2). In eq. (1.1),
$c_{i\alpha}$ denotes a spin ${1\over 2}$ fermion
operator, $P_G$ is the Gutzwiller projection, eliminating double occupation,
modelling on--site Coulomb repulsion and the second term  is an AF--
Heisenberg Hamiltonian, where the spin $\vec S_i$ is given by

$$
\vec S_i = c^\dagger_{i\alpha} {\vec\sigma_{\alpha\beta} \over 2}
c_{i\beta}. \eqno(1.2)
$$

The system described by the Hamiltonian (1.1) is called ``$t-J$ model".

According to [9], to match with the physics of high $T_c$ cuprates one
should take ${J\over t} \sim {1\over 3}$.

We analyse the $t-J$ model with a path--integral approach (see e.g. [10]).
We write the euclidean action, corresponding to the Hamiltonian (1.1),
with chemical potential $\mu$, at temperature $T$, in terms of spin
${1\over 2}$ (Grassmann) fermionic fields $\Psi_\alpha, \Psi^*_\alpha$. The
action is given by

$$
S(\Psi, \Psi^*) = \int^\beta_0 dx^0 \sum_i \Psi^*_{i\alpha} (\partial_0
+ \delta) \Psi_{i\alpha}
+\sum_{<ij>} \{-t (\Psi^*_{i\alpha} \Psi_{j\alpha} + h.c.)
$$
$$
- {J\over 2} |\Psi^*_{i\alpha} \Psi_{j\alpha} |^2\} + \sum_{i,j} u_{i,j}
\Psi^*_{i\alpha} \Psi^*_{j\beta} \Psi_{j\beta} \Psi_{i\alpha}, \eqno(1.3)
$$

where the two--body potential $u_{i,j}$ is given by

$$
u_{i,j} =\cases{+\infty & $i=j$ \cr
-{J\over 4} & $i,j$ \ n.n \cr} \eqno(1.4)
$$

and $\delta = \mu + J/2 $.
The last two terms in (1.3) correspond to a rewriting of the AF
Heisenberg term plus a hard--core repulsion replacing Gutzwiller projection.
[Summation over repeated spin indices is understood and dependence on
euclidean time $x^0$ is not explicitly exhibited; $\beta = {1\over
k_B T}$, with $k_B$ the Boltzmann constant].

For example, the grand canonical partition function of the $t-J$ model is
expressed in path--integral form as

$$
\Xi (\beta, \mu) = \int {\cal D} \Psi {\cal D} \Psi^* e^{-S
(\Psi, \Psi^*)} \eqno(1.5)
$$

\vfill\eject

\vskip 0.3truecm
{\sl 2. ``Chern--Simons representation" of the t--J model}
\vskip 0.3truecm

A key problem is to find a good Mean Field Approximation (MFA) to
analyze the $t-J$ model.

A priori, we can consider many possibilities as starting point for a MFA from
a general procedure valid in 2D: the ``Chern--Simons bosonization". Let
$W_\mu$ be a gauge field of gauge group $G$; define the (euclidean)
Chern--Simons action by

$$
S_{c.s.} (W) = {1\over 4\pi i} \int^\beta_0 dx^0 \int d\vec x
\epsilon^{\mu\nu\rho} {\rm tr} \ [W_\mu \partial _\nu W_\rho + {2\over 3}
W_\mu W_\nu
W_\rho] (x_0, \vec x); \eqno(2.1)
$$

denote by $S(\psi, \psi^*, W)$ the action obtained from $S(\psi, \psi^*)$
by minimally coupling $\psi, \psi^*$ to the gauge field $W$.

The output of ``Chern--Simons bosonization" can be summarised as follows:
In $2D$, for suitable choices [11] of the group $G$ and of real coefficient(s)
$k_G$, one can replace the path--integration over $\Psi, \Psi^*$ by
path--integration over $W$ and new spin ${1\over 2}$ fields, $\chi,
\chi^*$,
bosonic or fermionic depending on $\{G, k_G \}$, substituting the action
$S(\Psi, \Psi^*)$ by

$$
S(\chi, \chi^*, W) + k_G S_{c.s.} (W)\eqno(2.2)
$$

(with suitable b.c. [11]) and the new theory is exactly equivalent to the
original $t-J$ model.
\vskip 0.3truecm

\underbar{Remarks}

1) Although the C.S. bosonization is an exact identity if treated
without approximations, each one of these ``C.S. representations" in terms
of $\chi, \chi^*, W$ can be taken as a starting point for a different MFA.
As an axample, if we choose $G=U(1), k_{U(1)}=1$ and $\chi, \chi^*$
bosonic, one can reproduce [12] the ``slave--boson" approach [13].

2) A similar strategy works for the Fractional Quantum Hall Effect:
the system in a plateau  of Hall conductivity around a filling fraction
$\nu = {1\over 2\ell +1}, \ell$ integer, of the first Landau level appears
[14] to have a good MFA in terms of a ``C.S. bosonization" of the original
long--distance action with $G= U(1)$, $k_{U(1)} = 2\ell +1$.

\bigskip
Here we sketch how the C.S. bosonization works intuitively, while more
details are
given in the Appendix  (for a rather complete discussion see [11]). [In the
following, we usually omit explicit reference to the conjugate fields,
like $\Psi^*$].

Integrating out the time--component of the gauge field, $W_0$,
appearing linearly in the action, one obtains a constraint of the form

$$
j_0 (x) = {k_G \over 2\pi} \epsilon_{0\nu\rho} W^{\nu\rho} (x),
\eqno(2.3)
$$

where $W^{\mu\nu}$ is the field strength associated with
$W_\mu$ (in particular $W^{12}$ is the ``$G$--magnetic field") and $j_\mu$
is the ``$G$--current" of the matter field.

As a result a ``$G$--vortex" is attached to every particle described by
$\chi, \chi^*$, hence assigning to them a ``$G$--magnetic charge". These
particles
also carry a ``$G$--electric charge", since $\chi$ is minimally coupled to
$W_\mu$. The presence of both ``electric" and ``magnetic" charges implies a
Aharonov--Bohm (A--B) effect when these particles are exchanged, thus
introducing a ``phase" for every exchange [15].
If this ``A--B phase" is trivial
(+1), then the statistics of $\chi$ is unchanged after $W$--integration
and,
to match with that of $\Psi,$ the field $\chi$ must be taken fermionic.
If the ``A--B
phase" is --1, then integration over $W$ turns  the statistics
of a bosonic $\chi$ into fermionic, thus matching with that of $\Psi$ (after
$W$--integration).

Furthermore one can prove that the only effect of the C.S. term is to
introduce the ``A--B phase". These arguments give an idea why C.S.
bosonization is an exact identity.

To analyze the $t-J$ model we choose $\chi$ fermionic,

$$
G= U(1)\times SU(2), \quad k_{U(1)} =-2, \quad k_{SU(2)} =1. \eqno(2.4)
$$

The gauge field with gauge group $U(1)$ is denoted by $B_\mu$ and the gauge
field with gauge group $SU(2)$ is denoted by $V_\mu$. Why we made this
choice?

The basic argument is that with this choice, applying a dimensional
reduction $(2D \rightarrow 1D)$ and performing a MFA we are able [16] to
reproduce the exact results on the one--dimensional $t-J$ model
in the limit $t>> J$,
obtained by Bethe--Ansatz and Conformal Field Theory techniques, including
the ``semionic" (i.e. intermediate between bosonic and fermionic [15])
statistics  of spin and charge excitations.

\vskip 0.3truecm
{\sl 3. A formal separation of charge and spin degrees of freedom}
\vskip 0.3truecm

In $2D$ we separate
formally [17] the spin and charge degrees of freedom (d.o.f.)
of $\chi$ (the fermion of the $t-J$ model in the chosen ``$U(1) \times
SU(2)$ Chern--Simons representation") by a polar decomposition:

$$
\chi_\alpha = H \Sigma_\alpha. \eqno(3.1)
$$

In (3.1), $H$ is a spinless fermionic field (``holon") and it is
minimally coupled to
$B$ respecting a local $U(1)$ gauge invariance; $\Sigma_\alpha$ is a spin
${1\over 2}$ bosonic field (``spinon"), it is minimally coupled
to $V$ respecting a
local $SU(2)$ gauge invariance and it satisfies the constraint

$$
\Sigma^*_{\alpha j} \Sigma_{\alpha j} =1 \eqno(3.2)
$$

for each site $j$. An additional abelian local gauge invariance, called
here h/s, appears due to the decomposition (3.1) of $\chi$. It corresponds
to the addition of a local phase to $H$ and subtraction of the same phase
from $\Sigma_\alpha$.

\smallskip
\underbar{Remarks}

1) Fields with the same quantum numbers and constraints of
those appearing in the ``slave fermion" approach [18] are obtained from $H,
\Sigma_\alpha$ by performing a Holstein--Primakoff transformation [12], but
the
action derived in the $``U(1) \times SU(2)$ C.S. representation" differs
from the slave fermion one. At this stage it is explicitly given [17] by

$$
S (H, \Sigma, B, V)= \int^\beta_0 dx^0 \{\sum_j [H^*_j (\partial_0 - i
B_0(j) -\delta) H_j + i B_0 (j)
$$
$$
+ (1-H^*_j H_j) \Sigma^*_{j\alpha}
(\partial_0 + i V_0 (j))_{\alpha\beta} \Sigma_{j\beta}]
+ \sum_{<ij>} [(- t H^*_j e^{i\int_{<ij>} B} H_i \Sigma^*_{i\alpha} (P
e^{i\int_{<ij>} V} )_{\alpha\beta} \Sigma_{j\beta} +
$$
$$
+ h.c.) +{J\over 2} (1- H^*_j H_j) (1-H^*_i H_i) (|\Sigma^*_{i\alpha} (P
e^{i\int_{<ij>} V})_{\alpha\beta} \Sigma_{j\beta} |^2 -{1\over 2})]\}
- 2 S_{c.s.} (B) + S_{c.s.} (V). \eqno(3.3)
$$

2) The fermionic field $\Psi$ itself can be written as a product of
two $U(1) \times SU(2)$--gauge invariant fields, one constructed out of $H$
and $B$ and one out of $\Sigma_\alpha$ and $V$; the statistics of both these
fields is semionic, as originally suggested by Laughlin [19].

\bigskip

As usual in the path--integral formalism, to proceed we need a gauge--fixing
of the local gauge symmetries of the action [20].

We gauge fix the $U(1)$ symmetry by imposing a Coulomb gauge on $B$.
Integrating out the time component $B_0$, appearing linearly in the action,
one obtains

$$
B_\mu = \bar B_\mu + \delta B_\mu (H), \quad \mu = 1,2, \eqno(3.4)
$$

where $\bar B_\mu$ is a mean field introducing a flux $\pi$ per plaquette
$p$, i.e. $e^{i\int_{\partial p} \bar B} = -1$, and $\delta B_\mu$ is a
fluctuation term depending on holon density.

We gauge--fix the $SU(2)$ symmetry retaining the bipartite structure
appearing in the ground state at zero doping, i.e. the N\'eel state of the AF
Heisenberg model. For this purpose, we adopt the ``N\'eel gauge"  defined
by $\Sigma_{j\alpha} = \sigma_x^{|j|}$ $1\choose 0$, $|j|= j_1 + j_2$, where
$(j_1, j_2)$ denote the coordinates of the site $j$. The
``spins" $\Sigma^* \vec\sigma \Sigma$ are then forced in the N\'{e}el
configuration, leaving all the $SU(2)$ degrees of freedom in the gauge field
$V$, unconstrained. We decompose $V$ into a ``Coulomb component" $V^c$
satisfying

$$
\partial^\mu V_\mu^c =0 \quad \mu=1,2
$$

and gauge fluctuations around it, described by an $SU(2)$ field $g$.
Integrating out the time--component $V_0$, appearing linearly in the
action, we obtain an explicit expression of $V^c_\mu$ in terms of $g$ and
$H$, denoted by $V_\mu (g, H)$.

Following a strategy developed in $1D$ [16], with techniques patterned from a
proof of the diamagnetic inequality [21], we then find a $g$--configuration,
$g^m$, optimizing
the partition function of holons in a fixed $g$--background [17].

For small but non-vanishing doping concentration we find that $\bar V
(H) \equiv V^c (g^m, H)$ attaches to the holons a vortex of
$SU(2)$--vorticity $\pm \sigma_z {\pi\over 2}$, where $\sigma_z$ is the
diagonal Pauli matrix, and the sign depends on the N\'eel sublattice
where the holon is located. Furthermore we have $g^m_j
\sigma_x^{|j|} = \sigma_x^{|j|+1}$. We rewrite

$$
V^c_\mu (g, H) = \bar V_\mu (H)+ \delta V_\mu (g, H), \quad \mu=1,2, \eqno(3.5)
$$

where the fluctuation term $\delta V$ depends also on the spin d.o.f.
described by $g$, whereas $\bar V$ is independent of them.

\vskip 0.3truecm
\underbar{Remark}

At this stage, performing a simple field redefinition,
the action
of the $t-J$ model can be exactly written [17] in terms of $H$ and $g$
as $S = S_h + S_s$, where

$$
S_h = \int^\beta_0 dx^0 \{\sum_j H^*_j (\partial_0 - (\sigma_x^{|i|}
g^\dagger_j \partial_0 g_j \sigma_x^{|j|})_{11} - \delta) H_j
$$
$$
+ \sum_{<ij>}[- t H^*_j e^{-i \int_{<ij>} \bar B + \delta B} H_i
(\sigma^{|i|}_x g_i P (e^{i\int_{<ij>}\bar V +\delta V}) g_j
\sigma_x^{|i|})_{11} + h.c.] \}
$$
$$
S_s = \int^\beta_0 dx^0 \{\sum_j (\sigma_x^{|j|}g^\dagger_j \partial_0
g_j \sigma_x^{|j|})_{11} +
$$
$$
+\sum_{<ij>} {J\over 2} (1-H^*_i H_i) (1-H^*_j H_j) \bigl[|(\sigma_x^{|i|}
g_i^\dagger P(e^{i\int_{<ij>} \bar V + \delta V})g_j \sigma_x^{|j|})_{11}|^2 -
{1\over 2} \bigr]\} \eqno(3.6)
$$

We now make the first basic approximation neglecting $\delta B_\mu$ and
$\delta V_\mu$, i.e. the feed--back of charge fluctuations on $B$ and of
spin fluctuations on $V^c$, still retaining the holon dependence of $V^c$.
Presumably the main neglected effect is a statistic transmutation giving
rise to  semionic statistics  for holons and spinons, as in $1D$.

We argue that the statistics  here is less relevant than in $1D$ because we
expect in $2D$ the formation of a bound state with the quantum numbers of the
electron, due to gauge fluctuations discussed
later on.

\vskip 0.3truecm
{\sl 4. Low energy effective action for spin degrees of freedom}
\vskip 0.3truecm

To derive a low--energy effective action for the spin d.o.f., we first
rewrite $g$ in $CP^1$ form, introducing a spin ${1\over 2}$ field $b_\alpha$
through

$$
g_j = \pmatrix{b_{1j} & -b^*_{2j} \cr
b_{2j} & b^*_{1j} \cr} \eqno(4.1)
$$

with the constraint $b^*_{j\alpha} b_{j\alpha} =1$ for every site $j$.

Subsequently, we apply to $b_\alpha$ the standard treatment of AF systems
splitting it into an AF component, described by a spin ${1\over 2}$ complex
boson field $z_\alpha, \alpha =1,2$ and a ferromagnetic component which is
then
integrated out [22].

The continuum action is then given in $CP^1$ form by

$$
S(z, A) = \int_0^\beta dx^0 \int d\vec x \{|(\partial_0 - A_0) z_\alpha|^2
+ v_s^2|(\partial_\mu - A_\mu) z_\alpha|^2
+ \bar V^2 (H) z^*_\alpha z_\alpha \} (x^0\vec
x), \eqno(4.2)
$$

where $v_s$ is the ``spin velocity", $J$-dependent, and $A$ is the standard
``Hubbard--Stratonovich" gauge field of the $CP^1$ models. [In the
derivation we implicitly assumed that the $CP^1$ model is in the symmetric
phase; this is self--consistent with the behaviour discussed later on].

This action describes spin waves $\vec\Omega = z^* \vec\sigma z$ interacting
with the vortices appearing in $\bar V^2$, centered at the holon positions.
We evaluate approximately  the effect of $\bar V^2$ averaged over holon
positions at fixed density $\delta$. One can argue that this treatment can
be justified for $J << t$, because the holon appears to develop a large
effective mass, due to the coupling to soft spin fluctuations [23].

The output is an average

$$
\langle \bar V^2\rangle \sim -\delta \ln \delta. \eqno(4.3)
$$

Substituting in (4.2) $\bar V^2$  by $< \bar V^2>$  produces a mass for the
spin
waves, suggesting that the system should exhibit short--range AF, with a
correlation length $\xi_{AF} \sim (-\delta \ln \delta)^{-{1\over 2}} (\sim
\delta^{-{1\over 2}}$ for $\delta$ not too small).
This result is in agreement with data deduced from neutron experiments in
$La_{2-\delta}$$ Sr_\delta Cu O_4$, where a fit for $\xi_{AF}$ was
suggested [24] in terms of $\delta^{-{1\over 2}}$, as shown in fig. 3.

\bigskip
\underbar{Remark}

If $S(z,A)$ would be the full action, the spinons $z$ would be
logarithmically confined by the Coulomb interaction mediated by $A$, due to
the massive nature of $z$, but coupling with holons
yields deconfinement.

\vskip 0.3truecm
{\sl 5. Low energy effective action for charge degrees of freedom}
\vskip 0.3truecm
Introducing a flux $\pi$ per plaquette, the mean field $\bar B$ induces a
partition of the lattice in two N\'eel sublattices (here denoted $\uparrow$
and $\downarrow$) and it converts the spinless fermion $H$ into
2--components Dirac--like fermions of two--species:

$$
\psi^{(1)} = {\psi^{(i)}_\uparrow \choose \psi^{(1)}_\downarrow} \qquad
\psi^{(2)} =
{\psi^{(2)}_\downarrow\choose \psi^{(2)}_\uparrow}, \eqno(5.1)
$$

each component of them being supported on a N\'eel sublattice, as indicated
by the subscript.  The vertices of the cones of the ``Dirac"
energy--momentuum dispersion relation are centered at the four points $(\pm
{\pi \over 2}, \pm {\pi \over 2})$ in the Brillouin zone. (This phenomenon
is standard in the flux phase; see [22,25]). Up to a short--range term,
the continuum action for these fermions is given by

$$
S(\psi, A) = \int^\beta_0 dx^0 \int d\vec x \{\sum^2_{r=1} \bar\psi^{(r)}
[\gamma_0 (\partial_0 - \delta - e^{(r)} A_0)
- \tilde t (\parz - e^{(r)} \A)] \psi^{(r)}\} (x^0, \vec x), \eqno(5.2)
$$

where $\tilde t$ is a renormalized hopping parameter and $e^{(1)} =
\pmatrix{1 & 0 \cr
0 & -1 \cr}, e^{(2)} = \pmatrix{-1 & 0 \cr
0 & 1\cr}. $ [We adopt the standard notation

$$
\gamma_0 = \sigma_z \quad \gamma_\mu = (\sigma_y, \sigma_x) \quad \A =
\gamma_\mu A_\mu \quad \bar\psi = \psi^* \gamma_0 \quad].
$$

The components of the fermion fields in (5.2) supported in the two N\'eel
sublattices have opposite charge with respect to $A$, which is the same gauge
field appearing in (4.2).
It can be traced back as the gauge field of h/s gauge invariance.

With respect to the action discussed in [25], a crucial difference is the
appearance of the $\gamma_0 \delta$ term. Neglecting at first $A$, this
term produces a finite Fermi surface for the gapless components
$(\psi^{(1)}_
\uparrow, \psi^{(2)}_\downarrow$). The other two components have a gap and
they mix with the gapless ones, due to the presence of the (non--diagonal)
Dirac $\gamma$ matrices.

This mixing is expected to produce a reduction of the spectral weight in
the outer part in the reduced Brillouin zone scheme.

In a similar lattice model (with a twist of statistics between holon and
spinons) the shape of the Fermi surface for the ``electron" deduced in mean
fields is half--pocket like [26], showing a qualitative agreement with the
F.S.
deduced from ARPES experiments in underdoped cuprates [6] (see fig. 4).

\vskip 0.3truecm
{\sl 6. Final comments}
\vskip 0.3truecm

We can summarize the large distance behaviour of the  model in $U(1)
\times SU(2)$ C.S. representation (with $\delta B = \delta V =0)$, in terms
of a system of spinons, $z_\alpha$, whose dynamics is described by a $CP^1$
model with mass term $m\sim \delta^{1\over 2}$ (the main novelty) and a
system of Fermi liquid holons (obtained by integrating out the gapful Dirac
modes) with $\epsilon_F \sim t\delta$, interacting via an abelian
gauge field $A$.

The $U(1)$ effective action for $A$ then exhibits the basic features, such
as the Reizer [27] singularity in the propagator of the transverse
component of
$A$ and the existence of a characteristic energy scale for spinons, that
permits one to apply the ideas of [28] to deduce a linear in $T$ in--plane
resistivity for certain temperature range.

A more careful study of in--plane resistivity in the present model
is in prgress [29].

The analysis presented in this paper applies only to small doping
concentration
$\delta$, since we derived under this condition (see eq. 3.5--4.3)
the form of $\bar V$ used in the treatment, and
crucial to get a mass term for spinons.  Hence
the above description is applicable to underdoped materials, as explicitly
stated in Sec.1. The extension to other regions of the phase
diagram is  an open problem for further investigation.

\vskip 0.3truecm
{\sl Appendix: Chern--Simons bosonization}
\vskip 0.3truecm

We start by rewriting the grand canonical partition function $\Xi
(\beta,\mu)$ of the $t-J$ model in terms of the canonical partition function
$Z_N$ at fixed number of fermions $N$:

$$
\Xi (\beta, \mu) = \sum_N {e^{\beta \mu N} \over N!} Z_N (\beta) .
\eqno(A.1)
$$

Following a standard treatment of Feynman path--integral in the continuum
[10], adapted to the lattice, we can express $Z_N$ as a sum over virtual
trajectories of   fermions on a lattice with imaginary time. These
trajectories start at the imaginary time $x^0 =0$ at a set of lattice sites
$\{j_1, ..., j_N\}$ and they end at the imaginary time $x^0 = \beta$ at the
same set of sites permuted, $\{j_{\sigma (1)}, ..., j_{\sigma(N)}\}$,
where $\sigma$ is a
permutation (see fig.5). Each term in the sum has a weighting factor
$(-1)^{\epsilon(\sigma)}$,
where $\epsilon (\sigma)$ is the number of exchanges in
$\sigma$. By omitting this factor one reproduces the canonical partition
function, $Z^b_N$, of a fictitious ``bosonic $t-J$ model" obtained by
substituting fermions with bosons in the $t-J$ model. The hard--core
constraint forbids any intersection among the trajectories of the $N$
fermions and, by periodicity in imaginary time (the planes $x^0 =0$ and
$x^0 =\beta$ are identified, see [10]), every set of virtual trajectories
defines a link ${\cal L}$ (see fig.5). The factor $(-1)^{\epsilon(\sigma)}$
is a
topological invariant associated to that link, i.e. it does not change under
an arbitrary deformation of the link performed without allowing
intersections. According to a general theory [30] every topological invariant
of links can be represented as the expectation value of a ``Wilson loop"
supported on the link, in a gauge theory with Chern--Simons
action $S_{c.s.}$.

More precisely, let $W_\mu$ denote a gauge field of gauge group $G$ and
$k_G$ a real coefficient. Then, for a suitable choice  of $G$ and $k_G$ (for
explicit conditions, see [11]), we have

$$
(-1)^{\epsilon(\sigma)} = \int {\cal D} W e^{-k_G S_{c.s.} (W)}
{\rm Tr} \  P (e^{i\int_{{\cal L}}W_\mu dx^\mu}), \eqno(A.2)
$$

where ${\cal L}$ is a link associated with a set of virtual trajectories
whose end points are obtained by a permutation $\sigma$ from the initial
points. In (A.2) the last factor is the ``Wilson loop", where Tr denotes
the normalised trace and $P(\cdot)$ denotes the normal ordering, which
amounts to the usual time ordering $T(\cdot)$ for a fictitious ``time"
parametrising the link. Furthermore, the normalization is chosen such that
$\int {\cal D} W {\rm exp} \{-k_G S_{c.s.} (W)\} =1$.

Let us denote by $Z^b_N (W)$ the canonical partition function of the
modified ``bosonic $t-J$ model" with the boson field minimally
coupled to the gauge
field $W$. It is easy to verify (see e.g.[30] for the abelian and [12]
for the non-abelian case) that the dependence of $W$
in $Z^b_N (W)$ appears in the form of linear combinations of Wilson loops
on the links associated with virtual trajectories. Therefore the
identity $(A.2)$ implies that

$$
Z_N = \int {\cal D} W e^{-k_G S_{c.s.} (W)} Z^b_N (W). \eqno(A.3)
$$

Equation (A.3) is a ``bosonization formula" and it holds for all the
couples of $G$ and $k_G$ satisfying (A.2).
If we plug (A.3) into (A.1), we obtain

$$
\Xi = \int {\cal D} W e^{-k_G S_{c.s.} (W)} \Xi^b (W),\eqno(A.4)
$$

where $\Xi^b (W)$ is the grand--canonical partition corresponding to $Z^b_N
(W)$, given by

$$
\Xi^b (W) = \int {\cal D} \chi {\cal D}\chi^* e^{-k_G S (\chi, \chi^*,
W)}, \eqno(A.5)
$$

with $\chi$ a bosonic field. Combining
equations (A.4) and  (A.5), one obtains
the ``Chern--Simons bosonization" discussed in the text. It
is obvious that introducing further Chern--Simons gauge fields we can turn
$\chi$ into a fermion, following a similar procedure.

\vskip 0.3truecm
\underbar{Remarks}

1) A warning: in the above discussion subtle points like boundary
condition and ``framing" are completely ignored, see [11,30].

2) To have a more concrete feeling about formula (A.2), let us consider the
simple case $G=U (1), k_{U(1)} =1$, a choice for which the formula holds,
and, due to the
abelian nature of $G$, path--ordering is not needed, and the trace Tr
is trivial.

Let us denote by $\Sigma_\mu dx^\mu$ a ``singular $\delta$--like current"
supported on a surface $\Sigma$, whose boundary is given by the link ${\cal
L}$. Then one easily verifies that

$$
e^{i\int_{\cal L} W_\mu dx^\mu} = e^{i\int W_\mu \epsilon^{\mu\nu\rho}
\partial_\nu \Sigma_\rho d^3 x}. \eqno(A.6)
$$

We now compute

$$
\int {\cal D} W e^{{i\over 4\pi} \int \epsilon^{\mu\nu\rho} W_\mu
\partial_\nu W_\rho d^3 x} e^{i \int W_\mu
\epsilon^{\mu\nu\rho} \partial_\nu \Sigma_\rho d^3 x} \eqno(A.7)
$$

by a change of variable. Shifting  $W_\mu \rightarrow W_\mu + 2\pi
\Sigma_\mu$ and ``completing the square", one obtains

$$
\int{\cal D} W e^{{i\over 4\pi} \int \epsilon^{\mu\nu\rho}
W_\mu \partial_\nu W_\rho d^3 x} e^{-{i\over 2\pi} {1\over 2}
(2\pi)^2 \int e^{\mu\nu\rho} \Sigma_\mu\partial_\nu\Sigma_\rho d^3 x}
$$
$$
= e^{-i \pi \int \Sigma_\mu \epsilon^{\mu\nu\rho} \partial_\nu \sum_\rho d^3
x} = e^{-i\pi \int_{\cal L} \Sigma_\mu dx^\mu}. \eqno(A.8)
$$

The integral in the exponent of the last term in (A.8) receives a
contribution only when ${\cal L}$  crosses the surface $\Sigma$, i.e. for every
crossing appearing in ${\cal L}$, i.e. for every exchange of the particles
in the virtual trajectories, so that its value is given by $(-1)^
{\epsilon (\sigma)}$, q.e.d. .
\vskip 0.5truecm

{\sl Acknowledgments.} I wish to thank J. Fr\"ohlich, Z-B. Su and L. Yu for
the joy of collaboration. Useful discussions with N.P. Ong and F.Toigo
are also gratefully acknowledged.

\vskip 0.5truecm
{\bf References}
\vskip 0.3truecm
\item{[1]} Y. Iye, in ``Physical Properties of High Temperature
Superconductors III", D.M. Ginsberg, ed., World Scientific 1992.

\item{[2]} C.H. Pennington, C.P. Slichter, in ``Physical
Properties of High Temperature Superconductors II". D.M. Ginsberg, ed.,
World Scientific 1990.

\item{[3]} Z.X. Shen, D.S. Dessau, Phys. Rep. {\bf 253}, 1 (1995).

\item{[4]} B. Batlogg et al., Physica {\bf C235}, 130 (1994).

\item{[5]} M. Tagikawa et al., Phys. Rev. {\bf B43}, 247 (1991); {\bf 49},
4158 (1994); T. Imai et al., Physica {\bf C162}, 169 (1989).

\item{[6]} D.S. Marshall et al., Phys. Rev. Lett. {\bf 76}, 4841 (1996); A.G.
Loeser et al., Science {\bf 273}, 325 (1996); H. Ding et al, Nature  (London)
{\bf 382}, 51 (1996).

\item{[7]} F.C. Zhang, T.M. Rice, Phys. Rev. {\bf B37}, 5594, (1998).

\item{[8]} P.W. Anderson, Science {\bf 235}, 1196 (1987).

\item{[9]} T.M. Rice, in ``Strongly Interacting Fermions and High Tc
Superconductivity" (Les Houches 1991) Elsevier 1995.

\item{[10]}See e.g. J.W. Negele, H. Orland: ``Quantum many particle
systems". Addison Wesley 1988.

\item{[11]} J. Fr\"ohlich, T. Kerler, P.A. Marchetti, Nucl. Phys. {\bf
B374}, 511 (1992).

\item{[12]} J. Fr\"ohlich, P.A. Marchetti, Phys. Rev. {\bf B46}, 6535
(1992).

\item{[13]} P.W. Anderson, in ``Frontiers and Borderlines in Many--Particle
Physics", edited by R.A. Broglia et al., North--Holland 1988; Z. Zou, P.W.
Anderson Phys. Rev. {\bf B37}, 5594 (1988).

\item{[14]} See e.g. S.C. Zhang, T.H. Hansson, S. Kivelson, Phys. Rev. Lett.
{\bf 62}, 82 (1989); J. Fr\"ohlich, U. M. Studer, E.Thiran, in ``Fluctuating
Geometries in Statistical Mechanics and Field Theory" (Les Houches 1994), F.
David et al., eds. 1996.

\item{[15]} ``Fractional Statistics and Anyon Superconductivity", F. Wilczek,
ed., World Scientific 1990.

\item{[16]} P.A. Marchetti, Z.B. Su, L. Yu, Nucl. Phys. {\bf B482} [FS], 731
(1996).

\item{[17]} P.A. Marchetti, Z.B. Su, L. Yu, Mod. Phys. Lett. {\bf B12}, 173
(1998); Phys. Rev. {\bf B58}, 5808 (1998).

\item{[18]} D. Yoshioka, J. Phys. Soc. Japan {\bf 58}, 1516 (1989) and
references therein.

\item{[19]} R.B. Laughlin, Science {\bf 242}, 525 (1988); Int. J. Mod.
Phys. {\bf B5}, 1507 (1991).

\item{[20]} See e.g. V.N. Popov: ``Functional integral in quantum field
theory and statistical physics", Dordrecht Reidel 1983.

\item{[21]} D. Brydges, J. Fr\"ohlich, E. Seiler, Ann. Phys. {\bf 121}, 227
(1979).

\item{[22]} E. Fradkin, ``Field Theories of Condensed Matter Systems",
Addison Wesley 1991.

\item{[23]} See e.g. L.Yu, Z.B. Su, Y. Li, Chin. J. Phys. (Taipei) {\bf
31}, 579 (1993).

\item{[24]} R.T. Birgeneau et al., Phys. Rev. {\bf B38}, 6614 (1988).

\item{[25]} N. Dorey, N.E. Mavromatos, Phys. Lett. {\bf B250}, 107 (1990).

\item{[26]} X. Dai, Z.B. Su, L. Yu, Phys. Rev. {\bf B56}, 5583 (1997)

\item{[27]} P.A. Lee, N. Nagaosa, Phys. Rev. {\bf B46}, 5621 (1992)

\item{[28]} L.B. Ioffe, P.B. Wiegmann Phys. Rev. Lett. {65}, 653 (1990).

\item{[29]} P.A. Marchetti,  J.H. Dai, Z.B. Su, L. Yu, in preparation.

\item{[30]} E. Witten, Commun. Math. Phys. {\bf 121}, 351 (1989).
\vfill\eject
\epsfysize=19.5cm\epsfbox{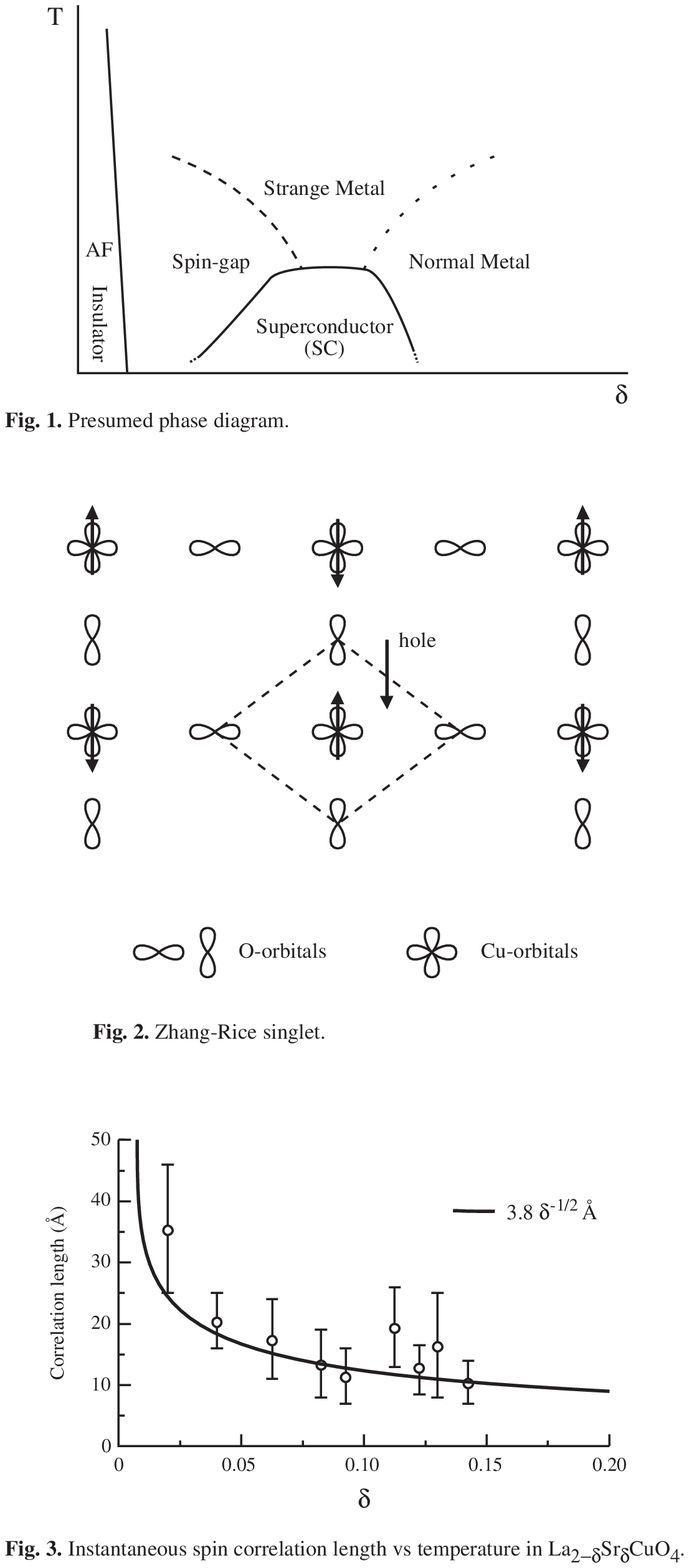}
\vfill\eject
\epsfysize=19.5cm\epsfbox{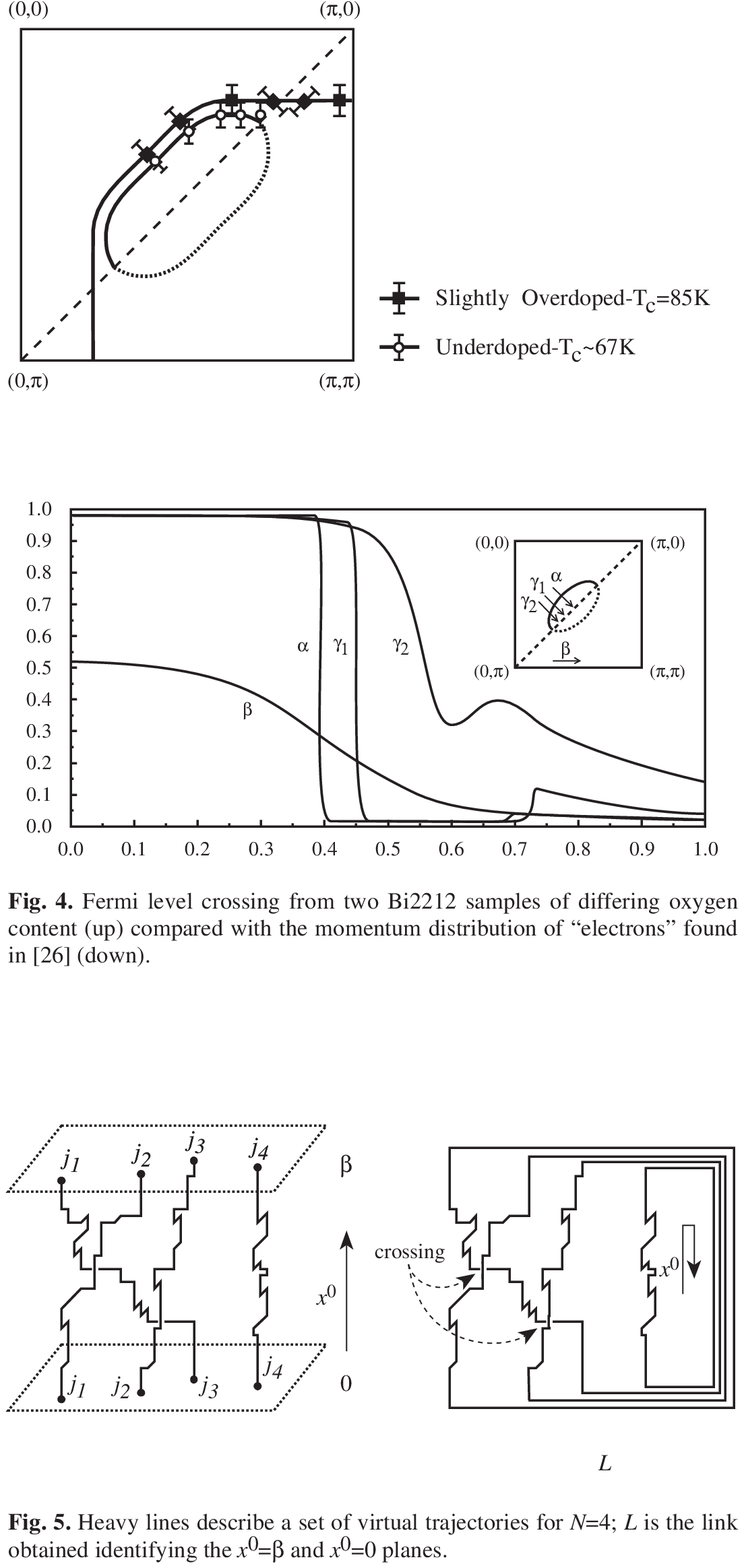}

\bye